\documentclass[%
 reprint,
superscriptaddress,
nofootinbib,
 amsmath,amssymb,
 aps,
]{revtex4-1}

\usepackage{graphicx}
\usepackage{dcolumn}
\usepackage{bm}

\begin{document}

\preprint{APS/123-QED}

\title{Innermost stable circular orbits of spinning test particles \\
in Schwarzschild and Kerr space-times}

\author{Paul I. Jefremov}
\email{paul.jefremow@zarm.uni-bremen.de}
\affiliation{ZARM - Center of Applied Space Technology and Microgravity, University of Bremen, Am Fallturm, 28359 Bremen, Germany}
\affiliation{Space Research Institute of Russian Academy of Sciences, Profsoyuznaya 84/32, Moscow 117997, Russia}

\author{Oleg Yu. Tsupko}
\email{tsupko@iki.rssi.ru}
\affiliation{Space Research Institute of Russian Academy of Sciences, Profsoyuznaya 84/32, Moscow 117997, Russia}
\affiliation{National Research Nuclear University MEPhI (Moscow Engineering Physics Institute), Kashirskoe Shosse 31, Moscow 115409, Russia}%

\author{Gennady S. Bisnovatyi-Kogan}
\email{gkogan@iki.rssi.ru}
\affiliation{Space Research Institute of Russian Academy of Sciences, Profsoyuznaya 84/32, Moscow 117997, Russia}
\affiliation{National Research Nuclear University MEPhI (Moscow Engineering Physics Institute), Kashirskoe Shosse 31, Moscow 115409, Russia}%

\date{\today}

\begin{abstract}
We consider the motion of classical spinning test particles in Schwarzschild and Kerr metrics and investigate innermost stable circular orbits (ISCO). The main goal of this work is to find analytically the small-spin corrections for the parameters of ISCO (radius, total angular momentum, energy, orbital angular frequency) of spinning test particles in the case of vectors of black hole spin, particle spin and orbital angular momentum being collinear to each other. We analytically derive the small-spin linear corrections for arbitrary Kerr parameter $a$. The cases of Schwarzschild, slowly rotating and extreme Kerr black hole are considered in details. For a slowly rotating black hole the ISCO parameters are obtained up to quadratic in $a$ and particle's spin $s$ terms. From the formulae obtained it is seen that the spin-orbital coupling has attractive character when spin and angular momentum are parallel and repulsive when they are antiparallel. For the case of the extreme Kerr black hole with co-rotating particle we succeed to find the exact analytical solution for the limiting ISCO parameters for arbitrary spin. It has been shown that the limiting values of ISCO radius and frequency do not depend on the particle's spin while values of energy and total angular momentum depend on it. We have also considered circular orbits of arbitrary radius and have found small-spin linear corrections for the total angular momentum and energy at given radius. System of equations for numerical calculation of ISCO parameters for arbitrary $a$ and $s$ is also explicitly written.
\begin{description}
\item[PACS numbers] 04.20.-q -- 04.25.-g


\end{description}
\pacs{?????? - ??????}
\end{abstract}

\pacs{?????? - ??????}
\maketitle


\section{Introduction}
In General Relativity the presence of a rotation (intrinsic angular momentum, or spin) of the central body influences motion of a particle orbiting it. Due to this reason the orbits of test particles differ in the Schwarzschild and Kerr backgrounds. When, in turn, a test particle has spin as well, it will also influence the particle's orbit. In particular, the motion of a spinning particle will differ from the non-spinning one even in the Schwarzschild background.

Let us consider a massive non-spinning test particle orbiting a central black hole (BH). There exists a minimal radius at which stable circular motion is still possible, it defines the so-called innermost stable circular orbit (ISCO) in given background. For the Schwarzschild background the radius of ISCO equals to $6M$\footnote{In this paper we use the system of units where $G=c=1$, the Schwarzschild radius $R_S=2M$, and other physical quantities which will be introduced further have the following dimensionalities: $[L]=[M]$, $[J]=[M]$, $[E]=1$, $[a]=[M]$, $[s]=[M]$.} \cite{Kaplan, LL2}. In the Kerr space-time circular motion is possible only in the equatorial plane of BH and the radius of ISCO depends on the direction of motion of the particle in comparison with the direction of BH rotation, whether they co-rotate or counter-rotate. Co-rotation and counter-rotation cases correspond to parallel and antiparallel orientation of vectors of the orbital angular momentum of the particle and the BH angular momentum. For the case of the extreme Kerr background the difference between these two variants is quite considerable: we have $9M$ for the antiparallel and $M$ for the parallel orientation \cite{Ruffini1971, Bardeen1972, LL2}. 

\begin{figure}
\centerline{\hbox{\includegraphics[width=0.5\textwidth]{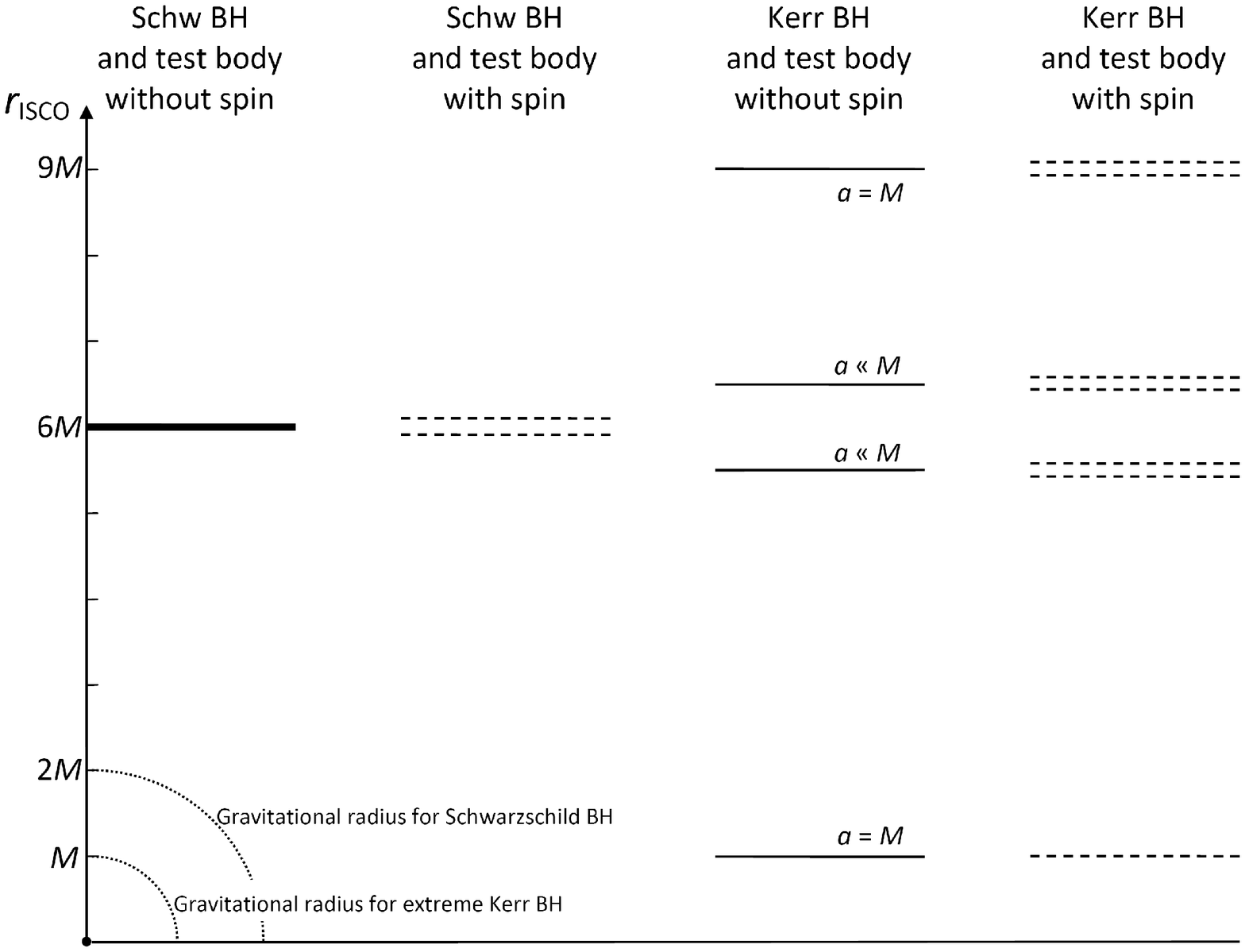}}}
\caption{The influence of BH and test-body's spin on the radii of ISCO. \\
In case of Schwarzschild BH the ISCO radius for spinless particles equals to $6M$, see thick solid line. For a particle with spin, ISCO radius splits in two cases, with different mutual orientation of spin and orbital angular momentum (parallel or antiparallel), see dash lines. \\
In the case of Kerr BH and spinless particle, ISCO radius splits in two cases, with co- and counter-rotation to BH (angular momentum is parallel or antiparallel to BH spin), see thin solid lines. For a particle with spin each line additionally splits in two dashed lines corresponding to different spin orientation. The only exception is the case of extreme Kerr BH with a co-rotating particle: in this case the ISCO radius is equal to $M$ and does not depend on magnitude and direction of particle spin.}
\label{figure-Im1}
\end{figure}

Values of the ISCO parameters (radius, total angular momentum, energy, orbital angular frequency) are determined by the Kerr parameter $a$. The main subject of the present paper is to investigate how the ISCO parameters (at a given value of $a$) are changed if a test particle has spin (Fig. \ref{figure-Im1}). We consider motion of a spinning test particles in the Schwarzschild and Kerr metrics in the equatorial plane and analytically derive the corrections for the parameters of ISCO taking spin to be a small parameter. We restrict our consideration by the case when vectors of black hole spin, particle spin and orbital angular momentum are collinear.

For non-spinning particles moving in the Schwarzschild metric the ISCO parameters were found by Kaplan \cite{Kaplan}. The solution of Einstein field equations around a rotating black hole was found by Kerr \cite{Kerr1963}. In the following works of Carter \cite{Carter68} and de Felice \cite{Felice68} the geodesic motion of general type was studied for the Kerr and Kerr-Newman (charged BH) geometry. See also Wilkins \cite{Wilkins1972} for bound orbits in Kerr metric and Dymnikova \cite{Dymnikova1986} for review. The parameters of ISCO in Kerr space-time for a non-spinning particle were obtained in the works by Ruffini \& Wheeler \cite{Ruffini1971} and Bardeen \cite{Bardeen1972}.

The problem of the motion of a classical spinning test body in General Relativity was considered in papers of Mathisson \cite{Mathisson1937}, Papapetrou \cite{Papapetrou1951a} and Dixon \cite{Dixon1970a, Dixon1970b, Dixon1978}, using different techniques. The equations of motion of a spinning test particle in a given gravitational field were derived in different forms; they are now referred to as Mathisson-Papapetrou-Dixon equations. From these equations it follows that the motion of the centre of mass and the particle rotation are connected to each other, and when the particle has spin the orbits will differ from geodesics of a spinless massive particle. In this paper we use the equations of motion in the form derived in Dixon's papers.

Influence of spin on orbits in the Schwarzschild metric was investigated in the paper of Corinaldesi and Papapetrou \cite{Papapetrou1951b}, and in the paper of Micoulaut \cite{Micoulaut1967}. A motion of a spinning test particle in Kerr metric was considered by Rasband \cite{Rasband1973}. Using integrals of motion arising from the symmetries in the Kerr space-time Rasband \cite{Rasband1973} derived the equations of motion of the radial coordinate of a spinning particle for the motion in the equatorial plane of a BH. Also in that work the spin-induced corrections for the radii of last stable orbits for Schwarzschild ($a=0$) and extreme Kerr ($a=M$) metric were illustrated through numerical calculations. In the paper by Tod, de Felice \& Calvani \cite{Tod1976} the influence of test body's spin on the radii of ISCO for different values of Kerr parameter $a$ including naked singularities was numerically calculated, see also papers 
of Abramowicz \& Calvani \cite{Abramowicz1979} and Calvani \cite{Calvani1980}. More general case of Kerr-Newman metric was analysed by Hojman \& Hojman \cite{Hojman1977}. Subsequently the number of works on this subject were published \cite{Suzuki1997, Suzuki1998, SaijoMaeda1998, TanakaMino1996, Apostolatos1996, Semerak1999, Semerak2007, Plyatsko2012a, Plyatsko2012b, Plyatsko2013, Bini2004a, Bini2004b, Bini2011a, Bini2011b, BiniDamour2014, Damour2008, Faye2006, Favata2011, Steinhoff2011, Steinhoff2012, Hackmann2014, Putten1999}. The detailed derivation of the equations of motion using the integrals of motion was presented in the work of Suzuki \& Maeda \cite{Suzuki1997} for the Schwarzschild case and in the work of Saijo \textit{et al.} \cite{SaijoMaeda1998} for the Kerr background. Tanaka \textit{et al.} \cite{TanakaMino1996} suggested that the radius of the last stable orbit is independent of the particle's spin in the extreme Kerr background (for co-rotating orbits).

Many authors used an effective radial potential for investigation of spinning particle orbits \cite{Rasband1973}, \cite{Tod1976}, \cite{Hojman1977, Suzuki1997, Suzuki1998, SaijoMaeda1998}, \cite{Favata2011}, \cite{Steinhoff2012}, \cite{Hackmann2014}. Method of calculation of ISCO parameters of spinning particle moving in Kerr metric is presented in details in the paper \cite{Favata2011}. The equations for the circular orbits in the equatorial plane resulting from the Mathisson-Papapetrou-Dixon equations appear to be irresolvable analytically, and ISCO parameters for arbitrary value of $a$ are supposed to be found numerically. Linear corrections in spin for the ISCO parameters in Schwarzschild metric were found by Favata \cite{Favata2011}. In this work we analytically obtain the small spin corrections for the ISCO parameters for the Kerr metric at arbitrary value of $a$.

For the extreme $a=M$ and almost extreme $a=(1-\delta)M$ Kerr BH we succeed to find the exact analytical solution for the ISCO parameters for arbitrary spin, with only restrictions connected with applicability of Mathisson-Papapetrou-Dixon equations. It has been shown that the limiting values of ISCO radius and frequency for $a=M$ do not depend on the particle's spin while values of energy and total angular momentum do depend on it.

The present paper is organised as follows. In section \ref{section-spinlessEP} we describe a motion of a test body without spin and introduce the notion of the effective potential (EP). In section \ref{section-ISCOspinless} we describe how to find ISCO parameters with the help of EPs, in the Schwarzschild and Kerr metric. In section \ref{section-MPD} we present the basic equations for motion of a spinning test body. In section \ref{section-EPspin} we introduce EPs for spinning particles and write explicitly equations for ISCO parameters for arbitrary Kerr parameter and arbitrary spin. In section \ref{section-arbitrary} we derive formulae for calculation of the small-spin linear corrections to ISCO parameters for arbitrary value of $a$. In sections \ref{section-schw}, \ref{section-slowly}, \ref{section-extreme-Kerr} we consider in details the cases of Schwarzschild metric, Kerr metric with $a \ll M$ and extreme Kerr metric. In section \ref{section-exact} we present an exact solution for extreme Kerr BH in the case of co-rotation. In section \ref{section-circular} we consider circular orbits of arbitrary radius. Section \ref{section-conclusions} is conclusions.

\section{Motion of a spinless test body. Effective potential}
\label{section-spinlessEP}

In the problems of motion of a test particle in a central field it is convenient to express the equations of motion for the radial coordinate in terms of a so-called "effective potential" energy.

In the Newtonian dynamics these equations for a particle moving in the gravitational field of a central body with mass $M$ can be written as (see, for example, \cite{Hobson, LL1, MTW})
\begin{equation}
\frac{1}{2} \left( \frac{d r}{dt} \right)^2 + U_N(r) = \varepsilon,
\label{e}
\end{equation}
where $\varepsilon$ is the total non-relativistic energy per unit mass, $U_N$ is the so-called effective potential (EP)
\begin{equation}
U_N  =  - \frac{GM}{r} +  \frac{L^2}{2r^2}.
\label{a}
\end{equation}
Here $L$ is the angular momentum per unit mass, $G$ is the gravitational constant.

In the framework of General Relativity, the Schwarzschild space-time is given by the expression for the interval (in spherical Schwarzschild coordinates $\{t,r,\theta,\varphi\}$) as:
\begin{equation}
\begin{split}
ds^{2} = & -\left(1-\frac{2M}{r}\right)dt^2 + \left(1-\frac{2M}{r}\right)^{-1}dr^2 +  \\
&+  r^2 \left(  d\theta^2 + \sin^2\theta \ d \varphi^2\right).
\end{split}
\label{d}
\end{equation}
In this space-time the equation of motion for the radial coordinate is \cite{MTW}
\begin{equation}
\left( \frac{d r}{d \tau} \right)^2= E^2 - \left( 1 -\frac{2M}{r} \right) \left( 1 +\frac{L^2}{r^2} \right),
\label{b}
\end{equation}
with the connection between $dt$ and $d\tau$ as
\begin{equation}
\frac{dt}{d\tau} = \frac{E}{1 - 2M/r} .
\end{equation}
Here $E$ is the total energy at infinity per unit particle rest mass, $L$ is the angular momentum per unit particle rest mass, $\tau$ is the proper time. This equation can also be presented in terms of an effective potential
\begin{equation}
\left( \frac{d r}{d \tau} \right)^2= E^2 - U_{Schw}^2.
\label{b1}
\end{equation}
Here we identify with the effective potential the following expression:
\begin{equation}
U_{Schw}= \sqrt{\left( 1 -\frac{2M}{r} \right) \left( 1 +\frac{L^2}{r^2} \right)}.
\label{c}
\end{equation}
Comparing equations (\ref{e}) and (\ref{b}) we should keep in mind that non-relativistic total energy $\varepsilon$ is not an analogue of $E$, since unlike the latter, it does not contain rest mass energy. Newtonian non-relativistic energy and effective potential are obtained as $\varepsilon = (E^2-1)/2$ and $U_N = (U_{Schw}^2-1)/2$, see \cite{MTW}.

Let us consider the effective potential in the Schwarzschild space-time (\ref{c}) as a function of radial coordinate. Then, the angular momentum $L$ plays a role of a parameter defining the shape of $U_{Schw}(r)$ curve. In the case $L>2 \sqrt{3} M$ the effective potential has two extrema: maximum and minimum, at the radii of which unstable and stable circular motion are possible correspondingly. In the case when $L$ equals to the boundary value $2 \sqrt{3} M$, two extrema of EP merge into one inflection point. This boundary value of $L$ defines parameters of the last stable orbit which is also called the innermost stable circular orbit (ISCO), i.\,e. the boundary orbit on which the finite motion is still possible. Further, when the angular momentum $L$ of a test body is less than $2\sqrt{3} M$, the EP does not have an extremum. For these values of angular momentum neither type of finite motion is possible and a test body will inevitably fall in the black hole whatever values of $E$ it may have.

In Boyer-Lindquist coordinates the Kerr metric is given by the expression \cite{LL2, Hobson}
\begin{equation}
\begin{split}
ds^2 & = -\left( 1 -\frac{2M r}{\Sigma} \right) dt^2 -\frac{4M a r \sin^2 \theta}{\Sigma} dt \ d\varphi +\frac{\Sigma}{\Delta}dr^2 \\
&+ \Sigma \ d\theta^2 + \left( r^2 +a^2  +\frac{2M ra^2 \sin^2\theta}{\Sigma} \right)\sin^2\theta \ d\varphi^2,
\end{split}
\label{}
\end{equation}
where $a$ is the specific angular momentum of a black hole, $\Sigma \equiv r^2 + a^2 \cos^2\theta$, $\Delta \equiv r^2 - 2 Mr + a^2$.

The equation of motion for the radial coordinate of a test particle in the equatorial plane ($\theta=\pi/2$) of Kerr metric is written as \cite{Hobson}
\begin{equation}
\begin{split}
& \left( \frac{dr}{d \tau} \right)^2= E^2 - \\
& -  \left[1 -\frac{2M}{r} -\frac{a^2(E^2 -1) -L^2}{r^2}   -\frac{2M (L -aE)^2}{r^3} \right],
\end{split}
\label{f}
\end{equation}
with the connection between $dt$ and $d\tau$
\begin{equation}
\frac{dt}{d\tau} = - \frac{2Ma}{r \Delta} L + \frac{E}{\Delta} \left( r^2 + a^2 + \frac{2Ma^2}{r} \right).
\end{equation}
Here $L \equiv L_z$ is a $z$-component of the angular momentum per unit rest mass, equal to the total angular momentum per unit mass, when $z$-axis is parallel to the axis of the rotation of BH. So $L>0$ and $L<0$ corresponds to the co-rotation and the counter-rotation cases respectively, and $a$ is always positive.

We see at the right-hand side of the equation (\ref{f}) that the expression in brackets is dependent on the energy $E$. For this reason the effective potential in the Kerr space-time cannot be defined in such a simple way as for the Schwarzschild space-time.

Let us write the equation for the radial motion as
\begin{equation}
\begin{split}
& r^4 \left( \frac{dr}{d \tau} \right)^2  =   (r^4+a^2 r^2 + 2Mra^2) E^2 \, - \\
& - 4MrLaE  + 2ML^2r - a^2 r^2 - L^2 r^2 -  r^4  + 2Mr^3 .
\end{split}
\end{equation}

It is convenient to write this equation as \cite{MTW}
\begin{equation}
\left( \frac{dr}{d \tau} \right)^2 = \frac{1}{r^4} (\alpha E^2 - 2 \beta E + \gamma) \, ,
\label{f1}
\end{equation}
where $\alpha$, $\beta$ and $\gamma$ are defined as:
\begin{equation}
\begin{split}
&\alpha = (r^2 + a^2)^2 - \Delta a^2 > 0 \, , \\
&\beta = \left[ a \left(r^2 + a^2  \right) - \Delta a  \right]   L  = 2M r a L \, , \\
&\gamma = a^2 L^2 - \Delta (r^2 + L^2)   \, .
\end{split}
\end{equation}
Let us rewrite (\ref{f1}) as
\begin{equation}
\begin{split}
r^4 \left( \frac{dr}{d \tau} \right)^2 & = \alpha \left( E - \frac{\beta +  \sqrt{\beta^2 - \alpha \gamma} }{\alpha} \right)  \times \\
& \times \left(  E - \frac{\beta -  \sqrt{\beta^2 - \alpha \gamma} }{\alpha} \right) \, ,
\end{split}
\end{equation}
and define EP equal to
\begin{equation}
U_{Kerr} (r; L)  =  \frac{\beta +  \sqrt{\beta^2 - \alpha \gamma} }{\alpha} \, .
\label{U-Kerr}
\end{equation}
The positive square root must be taken here, see \cite{MTW}. This definition keeps analogy with the EP in the Schwarzschild space-time (and, therefore, with the Newtonian dynamics as well).

In some other works, however, \cite{Chandra, Bardeen1972, Hobson} the entire right-hand side of the equation (\ref{f}) is used instead of the effective potential $U_{Kerr}$, especially for the investigation of circular orbits and the ISCO parameters. Despite that it contains energy, some authors \cite{Bardeen1972, Hobson} keep calling it an 'effective potential', because it has some features of the EP important for the investigation of circular orbits (see next section).\\

\section{ISCO for non-spinning particles in the Schwarzschild and Kerr space-times} \label{section-ISCOspinless}

Let us describe how to find ISCO parameters and first consider the Schwarzschild metric.

For a circular motion we need two conditions to be satisfied simultaneously:

(i) the radial velocity should be equal to zero:
\begin{equation}
\frac{dr}{d\tau} = 0 \, ,
\label{circ-1}
\end{equation}
what corresponds to the equality
\begin{equation}
E=U_{Schw} \, .
\label{circ-schw-1}
\end{equation}

(ii) the acceleration of the radial coordinate should be absent:
\begin{equation}
\frac{d^2r}{d\tau^2} = 0 \, .
\label{circ-2}
\end{equation}
Differentiating (\ref{b1}) with respect to $\tau$ and dividing by $\dot{r}$, we obtain
\begin{equation}
2 \ddot{r} = \frac{d}{dr} \left[ E^2 - U_{Schw}^2 \right] \, ,
\label{accel}
\end{equation}
so the condition (\ref{circ-2}) corresponds to the relation
\begin{equation}
\frac{d}{dr} \left[ E^2 - U_{Schw}^2 \right] = 0 \, .
\end{equation}
For a given $E$ we have
\begin{equation}
\frac{dU_{Schw}^2}{dr} = 0 \, , \quad   U_{Schw}>0 \, .
\label{circ-schw-2}
\end{equation}

Solving the system of equations (\ref{circ-schw-1}) and (\ref{circ-schw-2}), we obtain the expressions for the radius of the circular orbit $r$ and for the particle energy $E$ on it \cite{Kaplan, LL2}, as a function of $L$ in the form
\begin{equation}
r=\frac{L^2}{2M}\left[ 1 \pm \sqrt{1-\frac{12M^2}{L^2}} \right],
\label{ff}
\end{equation}
\begin{equation}
E =L\sqrt{\frac{1}{Mr}}\left( 1 -\frac{2M}{r} \right).
\label{ff2}
\end{equation}
In eq. (\ref{ff}) the upper sign corresponds to the stable circular orbits in the minimum of EP and the lower one stands for the unstable circular orbits in its maximum. In eq. (\ref{ff2}) radius $r$ given by (\ref{ff}) should be substituted. From the expression (\ref{ff}) we see that for $L<2 \sqrt{3} M$ the radius becomes complex and we do not have a circular orbit at all. Mathematically this means that the EP with such values of $L$ does not have extrema.

In order to find the last stable orbit (ISCO) we need to find the last value of $L$ at which the EP still has extremum. The ISCO takes place when points of maximum and minimum of EP merge. Therefore we need the third condition:

(iii) the inflection point of the EP satisfies relation:
\begin{equation}
\frac{d^2 U_{Schw}^2}{dr^2} = 0 \, .
\label{circ-schw-3}
\end{equation}
So, in order to find the parameters of ISCO ($r$, $E$, $L$), we need to solve three equations (\ref{circ-schw-1}), (\ref{circ-schw-2}), (\ref{circ-schw-3}) simultaneously.

From (\ref{ff}) it is evident that the points of maximum and minimum of EP merge at the marginal angular momentum $L=2\sqrt{3}M$. Thus, the ISCO parameters in the Schwarzschild space-time are \cite{Kaplan}:
\begin{equation}
\begin{aligned}
r_{\mathrm{\, ISCO}}& = 6 M , \\
L_{\mathrm{\, ISCO}}& = 2\sqrt{3} M , \\
E_{\mathrm{\, ISCO}}& = \sqrt{\frac{8}{9}} \, .
\end{aligned}
\end{equation}

In the Kerr space-time the finding of ISCO parameters is more complicated. ISCO in the Kerr space-time were considered in \cite{Ruffini1971}, \cite{Bardeen1972}, where, in particular, parameters of ISCO for the extreme Kerr BH were found. This problem is described at length, for example, in the textbook by Hobson \textit{et al.} \cite{Hobson}.

It is convenient to solve the problem by introducing a function
\begin{equation}
V(r;L,E)=  \frac{1}{r^4} (\alpha E^2 -2\beta E +\gamma ) .
\label{V-spinless}
\end{equation}
The function $V(r;L,E)$ is the right-hand side of eq. (\ref{f1}) and has qualities which are important in our research. The expression for acceleration is given by the first derivative $\ddot r = (1/2) dV/dr$ (compare with (\ref{accel})), and stability of the circular orbit is given by the sign of its second derivative $d^2 V/d r^2$. It is convenient to use variable $u=1/r$ instead of $r$ and variable $x=L-aE$ instead of $L$.

For the circular motion we need the conditions for the velocity (\ref{circ-1}) and the acceleration (\ref{circ-2}) to be satisfied simultaneously. Therefore for the Kerr metric we have from (\ref{V-spinless}) the system of equations
\begin{equation}
\left\{
\begin{aligned}
V &= 0 , \\
\frac{dV}{dr} &=0 . \\
\end{aligned}
\right.
\label{system-r}
\end{equation}
The usage of $u$ instead of $r$ does not change anything in the form of the system: since $dV/dr=(dV/du)(du/dr)$, the condition $dV/dr=0$ is equivalent to $dV/du=0$.

The solution of the system (\ref{system-r}) defines the set of parameters $x=L-aE$ and $E$ for stable and unstable circular orbits as functions of $u$ \cite{Hobson}. For the stable circular orbit we have
\begin{equation}
\begin{aligned}
x(u) &=-\frac{a \sqrt{u}\pm \sqrt{M}}{\left[ u\left(1-3Mu\mp 2a \sqrt{Mu^3}\right)\right]^{1/2}},\\
E(u) &=\frac{1-2 M u \mp a \sqrt{M u^3}}{\left(1-3Mu\mp 2a \sqrt{Mu^3}\right)^{1/2}}.\\
\end{aligned}
\label{xE}
\end{equation}
Using $L=x+aE$, we have:
\begin{equation}
L(u) = \mp \frac{\sqrt{M}  (1+a^2 u^2 \pm  2a \sqrt{Mu^3}) }{\left[ u\left(1-3Mu\mp 2a \sqrt{Mu^3}\right)\right]^{1/2}} .
\label{L}
\end{equation}
The upper sign corresponds to the antiparallel orientation of particle's angular momentum $L$ and BH spin $a$ (counter-rotation), the lower -- to the parallel one (corotation).

In order to find the parameters of ISCO, we need to add the condition
\begin{equation}
\frac{d^2 V}{dr^2} = 0 .
\label{system-r-third}
\end{equation}
Written in terms of $u$ the third condition transforms to \cite{Hobson}:
\begin{equation}
\frac{d^2 V}{dr^2} = \frac{d^2 V}{du^2} \left(  \frac{du}{dr} \right)^2  +  \frac{dV}{du} \frac{d^2 u}{dr^2}  = u^3 \left(  \frac{d^2 V}{du^2}  + 2 \frac{dV}{du} \right) = 0  .
\end{equation}
Since $dV/du=0$ for a circular orbit, the third condition is $d^2 V/du^2=0$.

The three equations $V=0$, $dV/du=0$, $d^2V/du^2=0$ form a closed system on three parameters of ISCO $E$, $x$ and $u$, which are then dependent only on $M$ and the Kerr parameter $a$.

Solving this system \cite{Hobson}, we obtain the equation for the inverse ISCO radii in the Kerr metrics:
\begin{equation}
1- 3 a^2 u^2 - 6 u M \mp 8 a \sqrt{Mu^3} =0 ,
\label{u-0}
\end{equation}
or
\begin{equation}
r^2 - 6Mr - 3a^2 \mp 8a \sqrt{Mr} =0 .
\label{r}
\end{equation}
Analytic solution for the ISCO radius can be found in the paper \cite{Bardeen1972}. Solutions of (\ref{u-0}) should be substituted into (\ref{xE}) and (\ref{L}) for finding $E$ and $L$.

In the limit $a = 0$, we have $r_{\mathrm{\, ISCO}} = 6M$ in the Schwarzschild case. In the extreme Kerr limit $a = M$ we obtain
\begin{equation}
r_{\mathrm{\, ISCO}} = 9M , \; L_{\mathrm{\, ISCO}} = - \frac{22}{3\sqrt{3}} M , \;  E_{\mathrm{\, ISCO}} =  \frac{5}{3 \sqrt{3}}
\end{equation}
for the counter-rotating orbit and
\begin{equation}
r_{\mathrm{\, ISCO}} = M , \; L_{\mathrm{\, ISCO}} = \frac{2}{\sqrt{3}} M , \;  E_{\mathrm{\, ISCO}} =  \frac{1}{\sqrt{3}}
\end{equation}
for the co-rotating case \cite{Ruffini1971}, \cite{Bardeen1972}.\\

\section{Motion of a spinning test body in the equatorial plane of a BH} \label{section-MPD}

For a description of the influence of body's spin on the parameters of its circular orbits we use the Mathisson-Papapetrou-Dixon (MPD) \cite{Mathisson1937, Papapetrou1951a, Dixon1970a, SaijoMaeda1998} equations of motion:
\begin{equation}
\begin{split}
&\frac{Dp^\mu}{D\tau}=-\frac{1}{2}R^{\mu}{}_{\nu \rho \sigma}v^{\nu}S^{\rho \sigma} ,\\
&\frac{DS^{\mu \nu}}{D\tau}=p^\mu v^\nu - p^\nu v^\mu.
\label{g}
\end{split}
\end{equation}
Here $D/D \tau$ is a covariant derivative along the particle trajectory, $\tau$ is an affine parameter of the orbit \cite{SaijoMaeda1998}, $R^{\mu}{}_{\nu \rho \sigma}$ is the Riemannian tensor, $p^\mu$ and $v^\mu$ are 4-momentum and 4-velocity of a test body,  $S^{\rho \sigma}$ is its spin-tensor. The equations were derived under the assumption that characteristic radius of the spinning particle is much smaller than the curvature scale of a background spacetime \cite{SaijoMaeda1998} (see also \cite{Rasband1973}, \cite{Apostolatos1996}) and the mass of a spinning body is much less than that of BH.

These equations are, however, incomplete, because they do not define which point on the test body is used for spin and trajectory measurements, so we need some extra condition to do that \cite{Papapetrou1951b}. We use the condition of Tulczyjew \cite{Tulczyjew1959, Dixon1970a, SaijoMaeda1998} that fixes the centre of mass of the test body
\begin{equation}
p_\mu S^{\mu \nu}=0.
\label{}
\end{equation}

Let us consider the motion of a spinning particle in the equatorial plane of Kerr BH. In this case the angular momentum of a spinning particle is always perpendicular to the equatorial plane \cite{SaijoMaeda1998}. Therefore we can describe the test particle spin by only one constant $s$ which is the specific spin angular momentum of the particle. Value $|s|$ indicates the magnitude of the spin and $s$ itself is its projection on the $z$-axis. It is more evident to think of the spin in terms of the particle's spin angular momentum $\mathbf{S_1}=sm\mathbf{\hat{z}}$ which is parallel to the BH spin angular momentum $\mathbf{S_2}=aM\mathbf{\hat{z}}$, when $s>0$, and antiparallel, when $s<0$. Here $\mathbf{\hat{z}}$ is a unit vector in the direction of the $z$-axis and $m$ is a mass of the particle \cite{SaijoMaeda1998}, \cite{Favata2011}.

Saijo \textit{et al} \cite{SaijoMaeda1998} have derived the equations of motion of a spinning test particle for the equatorial plane of Kerr BH. The equations of motion for the variables $r$, $t$, $\varphi$ in this case have the form \cite{SaijoMaeda1998}
\begin{equation}
\begin{split}
& (\Sigma_s \Lambda_s \dot r)^2 =R_s,\\
& \Sigma_s \Lambda_s \dot t =a\left( 1 +\frac{3Ms^2}{r \Sigma_s}\right)\left[ J -(a+s)E \right] +\frac{r^2 +a^2}{\Delta}P_s,\\
& \Sigma_s \Lambda_s \dot \varphi =\left( 1 +\frac{3Ms^2}{r \Sigma_s}\right)\left[ J -(a+s)E \right] +\frac{a}{\Delta}P_s.\\
\end{split}
\label{spin-eqs}
\end{equation}
where
\begin{equation}
\begin{split}
& \Sigma_s= r^2 \left(1 -\frac{Ms^2}{r^3}\right),\\
& \Lambda_s= 1 - \frac{3 M s^2 r [-(a + s) E +J]^2}{\Sigma_s^3},\\
&R_s = P_s^2 - \Delta \left\{ \frac{\Sigma_s^2}{r^2} + [-(a + s) E +J]^2 \right\}, \\
&P_s = \left[r^2 + a^2 + \frac{a s (r + M)}{r} \right]  E  -   \left(a + \frac{M s}{r} \right) J.
\end{split}
\end{equation}
Here $\dot{x} \equiv dx/d \tau$ and the affine parameter $\tau$ is normalised as $p^{\nu} v_{\nu} =-m$ \cite{SaijoMaeda1998}; $E$ is the conserved energy per unit particle rest mass, and $J=J_z$ is the conserved total angular momentum per unit particle rest mass which is collinear to the spin of a BH.

Equation for radial motion can be rewritten as
\begin{equation}
(\Sigma_s \Lambda_s \dot r)^2 =\alpha_s E^2 -2\beta_s E +\gamma_s,
\label{l}
\end{equation}
where
\begin{equation}
\begin{split}
&\alpha_s = \left[r^2 + a^2 + \frac{a s (r + M)}{r} \right]^2 - \Delta (a + s)^2,\\
&\beta_s = \left[ \left(a + \frac{M s}{r} \right) \left(r^2 + a^2 + \frac{a s (r + M)}{r} \right) - \Delta (a + s) \right]J,\\
&\gamma_s = \left(a + \frac{M s}{r} \right)^2 J^2 - \Delta \left[r^2 \left(1 - \frac{M s^2}{r^3}\right)^2 + J^2 \right].
\end{split} \label{abc-spin}
\end{equation}
This form is used by Favata \cite{Favata2011}. Coefficients $\alpha_s$, $\beta_s$, $\gamma_s$ agree with Rasband \cite{Rasband1973} results.

\section{The effective potential and circular orbits for spinning test particles} \label{section-EPspin}

The question of what we should consider as the EP in the case when a particle has spin does not differ from the "spinless case" in the Kerr space-time. If we want to keep analogy with the EP in the Schwarzschild background, we should define it as the solution of equation $\alpha_s E^2 -2\beta_s E +\gamma_s = 0$ with $\alpha_s$, $\beta_s$, $\gamma_s$ from (\ref{abc-spin}).

EP in this form can be found in the paper \cite{SaijoMaeda1998}:
\begin{equation}
U_{sk}(r;J,s) =  \frac{\beta_s +  \sqrt{\beta_s^2 - \alpha_s \gamma_s} }{\alpha_s} \, .
\label{U-Kerr-spin}
\end{equation}
with $\alpha_s$, $\beta_s$, $\gamma_s$ from (\ref{abc-spin}), see also Favata \cite{Favata2011}, Rasband \cite{Rasband1973}, Suzuki \& Maeda \cite{Suzuki1997}, \cite{Suzuki1998}, Steinhoff \& Puetzfeld \cite{Steinhoff2012}. For a review of misprints and comparison of effective potentials presented in different works see Appendices of the paper \cite{Steinhoff2012}.

Using this EP for finding the ISCO parameters is presented, for example, in the paper of Favata \cite{Favata2011}. It is shown that conditions for a circular orbit (the vanishing of $\dot{r}$ and $\ddot{r}$) are equivalent to
\begin{equation}
E=U_{sk}, \quad \frac{dU_{sk}}{dr}=0 \, .
\end{equation}
For the ISCO the third condition is added:
\begin{equation}
\frac{d^2U_{sk}}{dr^2}=0 \, .
\end{equation}
A general solution of these three equations for the ISCO of a spinning particle is supposed to be found numerically. In the small-spin limit, Favata \cite{Favata2011} analytically had found parameters of ISCO in the Schwarzschild metric.

Since our purpose is to obtain the analytical expressions for circular orbits in the Kerr metric, it is more convenient to define the 'effective potential' similar to (\ref{V-spinless}), by analogy with the work \cite{Bardeen1972}. We rewrite equation (\ref{l}) as:
\begin{equation}
\dot{r}^2 =\frac{1}{\Sigma_s^2 \Lambda_s^2} (\alpha_s E^2 -2\beta_s E +\gamma_s) ,
\end{equation}
and define the effective potential as its right-hand side:
\begin{equation}
V_s(r;J,E)= \frac{1}{\Sigma_s^2 \Lambda_s^2} (\alpha_s E^2 -2\beta_s E +\gamma_s) .
\end{equation}
It can be easily shown that solution of system of equations determining circular orbits is not changed if the effective potential is multiplied by some function which is not equal to zero (with its first and second derivatives) for physically relevant parameter values. Therefore it is more convenient for analytical calculations to define effective potential as
\begin{equation}
V_s(r;J,E)= \frac{1}{r^4} (\alpha_s E^2 -2\beta_s E +\gamma_s) ,
\label{V-spin}
\end{equation}
with $\alpha_s$, $\beta_s$, $\gamma_s$ from (\ref{abc-spin}). Notice that in the case of a spinning particle we use the total angular momentum $J$ instead of the orbital angular momentum $L$.\footnote{For a spinless particle the conserved quantity is the orbital angular momentum $L_z$, whereas in the case of a spinning particle the conserved quantity is the total angular momentum $J_z$, which includes spin terms \cite{SaijoMaeda1998}. In this case 'orbital angular momentum' at infinity $L_z$ can also be introduced as $L_z=J_z - s$, see \cite{SaijoMaeda1998}. Further we can say that in both cases total angular momentum is conserved with a note that in spinless case it consists of orbital angular momentum part only.} Further, for the sake of convenience, we shall change variables and work not with $r$ and $J$ but with $u=1/r$ and $x=J-aE$, so the function $V_s(u; x,E)$ will be used.

\begin{figure}
\centerline{\hbox{\includegraphics[width=0.5\textwidth]{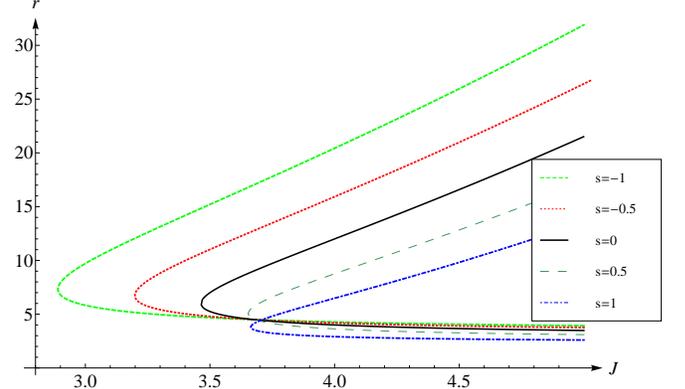}}}
\caption{Dependence of the radii of the stable (upper branches) and unstable (lower branches) circular orbits on the total angular momentum, for different values of spin $s$. All parameters are given in units of $M$, see also \cite{Rasband1973}.}
\label{figure-ImC}
\end{figure}

In order to find the circular orbits we need to solve the system of equations
\begin{equation}
\left\{
\begin{aligned}
V_s &= 0 \, ,\\
\frac{dV_s}{du} &=0 \, .\\
\end{aligned}
\right.
\label{n}
 \end{equation}
The explicit form of these equations is given in the first two equations in the system (\ref{p}). The solution of this system defines the set of parameters $x$ and $E$ for stable and unstable circular orbits as functions of $s$ and $u$. In Figure \ref{figure-ImC} we present the numerically calculated dependence of $r$ on total angular momentum $J$ and the influence of spin on it for the Schwarzschild background, see also \cite{Rasband1973}.

The point at which upper and lower branches on Figure \ref{figure-ImC} meet defines the ISCO for the given value of spin. In order to find the dependence of the radius of the ISCO on spin, we need to add the condition $d^2 V_s/dr^2= 0$, which is equivalent to

\begin{equation}
\begin{aligned}
&\frac{d^2 V_s}{du^2} = 0,\\
&\\
\end{aligned}
\label{n1}
\end{equation}
to the system (\ref{n}). The resulting system for finding the ISCO parameters, thus, consists of three equations (\ref{n}) and (\ref{n1}) and has the explicit form
\begin{widetext}
\begin{equation}
\left\{
\begin{aligned}
&(1 + 2 a s u^2 - s^2 u^2 + 2 M s^2 u^3) E^2 +  (-2 a u^2 x + 2 s u^2 x - 6 M s u^3 x - 2 a M s^2 u^5 x) E -1 + 2 M u - a^2 u^2 + 2 M s^2 u^3\\
&- 4 M^2 s^2 u^4 + 2 a^2 M s^2 u^5 - M^2 s^4 u^6 + 2 M^3 s^4 u^7 - a^2 M^2 s^4 u^8 - u^2 x^2 + 2 M u^3 x^2 + 2 a M s u^5 x^2 + M^2 s^2 u^6 x^2= 0 \, ,\\
&(4 a s u - 2 s^2 u + 6 M s^2 u^2) E^2 +(-4 a u x + 4 s u x - 18 M s u^2 x - 10 a M s^2 u^4 x) E +2 M - 2 a^2 u + 6 M s^2 u^2 - 16 M^2 s^2 u^3\\
& + 10 a^2 M s^2 u^4 - 6 M^2 s^4 u^5 + 14 M^3 s^4 u^6 - 8 a^2 M^2 s^4 u^7 - 2 u x^2 + 6 M u^2 x^2 + 10 a M s u^4 x^2 + 6 M^2 s^2 u^5 x^2=0 \, ,\\
&(4 a s - 2 s^2 + 12 M s^2 u) E^2 + (-4 a x + 4 s x - 36 M s u x - 40 a M s^2 u^3 x)E -2 (a^2 - 6 M s^2 u + 24 M^2 s^2 u^2 - 20 a^2 M s^2 u^3\\
&+ 15 M^2 s^4 u^4 - 42 M^3 s^4 u^5 + 28 a^2 M^2 s^4 u^6 + x^2 - 6 M u x^2 - 20 a M s u^3 x^2 - 15 M^2 s^2 u^4 x^2) =0 \, .
\end{aligned}
\right.
\label{p}
\end{equation}
\end{widetext}
These three equations form a closed system on three parameters of ISCO $E$, $x$ and $u$, which are then dependent only on the Kerr parameter $a$ and particle's spin $s$. This system can be used for numerical calculation of $r$, $E$, $J$ of ISCO at given $a$ and $s$.

Using EP in the form $U_{sk}(r;J)$ (\ref{U-Kerr-spin}) permits of splitting that system of three equations into two parts: equations $dU_{sk}/dr=0$ and $d^2U_{sk}/dr^2=0$ contain only $r$ and $J$, and can be solved independently from equation $E=U_{sk}$. So actually the problem is reduced to solving of the system of two equations. It is clearly seen in the case of the Schwarzschild metric for a spinless particle, see Section 3. In the case of using EP in the form $V_s(u; x,E)$ (\ref{V-spin}) the equations $dV_s/du=0$ and $d^2V_s/du^2=0$ contain not only $u$ and $x$, but also $E$, so that in order to find these three parameters it is necessary to solve simultaneously three equations. Disadvantage of $U_{sk}(r;J)$ is that it contains radicals, and each differentiation increases complexity of expressions. EP in the form $V_s(u; x,E)$, on the contrary, has polynomial form with respect to $u$, and each differentation makes the expressions simpler. All these properties are present in the spinless case also, and it is convenient to look for the ISCO parameters for the Schwarzschild case using $U_{Schw}(r;L)$ and the ISCO parameters for the Kerr case using $V_s(u; x,E)$.

Another important characteristics of the particle circular motion is its angular velocity. We calculate the spin corrections for the orbital angular frequency of the particle at the ISCO, as seen from an observer at infinity. This angular frequency is defined as
\begin{equation}
\Omega \equiv \frac{d \varphi / d\tau}{ dt / d\tau}.
\label{Omega-definition}
\end{equation}
The values $d \varphi / d\tau$ and $dt / d\tau$ are found from the second and the third equations in (\ref{spin-eqs}), where we should substitute values of $r$, $E$ and $J$ at a given orbit. To find the ISCO frequency $\Omega_{\mathrm{\, ISCO}}$ we need to use the ISCO values of $r$, $E$ and $J$, see \cite{Favata2011}.

\section{Small-spin linear corrections for ISCO parameters at arbitrary Kerr parameter} \label{section-arbitrary}

The system (\ref{p}) does not admit of an analytical solution for arbitrary values of $a$ and $s$. Using $s\ll M$, we can expand the equations of the system (\ref{p}) into series of powers of $s$.

In this section we find analytical solution for ISCO parameters with small-spin linear corrections, for arbitrary value of $a$.

To perform analytical calculations and get analytical results for small-spin corrections, it is more convenient to use the following equivalent system of equations instead of the system (\ref{n}) and (\ref{n1}):
\begin{equation}
\left\{
\begin{aligned}
\frac{1}{2} \frac{dV_s}{du} & = 0 \, ,\\
\frac{1}{2} \frac{dV_s}{du} \, u - V_s & =0 \, ,\\
\frac{1}{2} \frac{d^2V_s}{du^2} & =0 \, .\\
\end{aligned}
\right.
\label{new-system}
\end{equation}
Similar approach for spinless case is used in \cite{Hobson}.

We shall seek for solution of the system (\ref{new-system}) in the form
\begin{equation}
\begin{aligned}
x &=x_0+s x_1,\\
E &=E_0+ s E_1,\\
u &=u_0+ s u_1.\\
\end{aligned}
\label{podstanovka}
\end{equation}
Here $x_0$, $E_0$, $u_0$ are the values of $x$, $E$, $u$ in the spinless case, and $x_1$, $E_1$, $u_1$ are the linear corrections associated with spin. The dimensionalities are $[x]=[x_0]=[M]$, $[x_1]=1$, $[E]=[E_0]=1$, $[E_1]=[M^{-1}]$, $[u]=[u_0]=[M^{-1}]$, $[u_1]= [M^{-2}]$, $[s]=[M]$. Further we introduce these expressions into (\ref{new-system}) and linearise it in spin. Since the equations of the system are to be satisfied for arbitrary values of spin, we obtain the system of six equations for six unknowns $x_0, E_0, u_0, x_1, E_1, u_1$.

First, there are three equations for three unknowns $x_0$, $E_0$, $u_0$, at given $a$:
\begin{equation}
\left\{
\begin{aligned}
M - a^2 u_0 - x_0^2  u_0 - 2 a E_0 x_0 u_0    +   3 M x_0^2 u_0^2  = 0 \, ,\\
1 - E_0^2 - M u_0 + M x_0^2 u_0^3   =  0 \, ,\\
- a^2 - x_0^2   -2 a E_0 x_0  + 6 M x_0^2 u_0  =0 \, .\\
\end{aligned}
\right.
\label{system-null}
\end{equation}
These equations describe the parameters of ISCO for the spinless case. Solution of the first two equations in this system gives us the values of $x_0$ and $E_0$ for an arbitrary circular orbit of given inverse radius, see (\ref{xE}). Solving three equations simultaneously, we can obtain the equation for $u_0$, see (\ref{u-0}).

The solution of equation (\ref{u-0}) for $u_0$ has a complicated form. Therefore it is more convenient to express the six unknowns not as explicit functions of $a$ but as the explicit functions of $a$ and $u_0$, keeping in mind that $u_0$ can be found from eq.(\ref{u-0}) at arbitrary $a$. Before we proceed we need to notice that representation of all unknowns via $u_0$ given below could be rewritten in a different form using (\ref{u-0}).

To get $x_0$ and $E_0$ for ISCO we can substitute $u=u_0$ into the expressions for an arbitrary circular orbit (\ref{xE}). But it is more convenient to obtain directly the connections between parameters $x_0$, $E_0$, $u_0$ for ISCO from three equations (\ref{system-null}). It is easy to obtain that:
\begin{equation} \label{x0E0viau0}
x_0^2 = \frac{1}{3 u_0^2}, \quad E_0^2 = 1 - \frac{2}{3} M u_0 \, .
\end{equation}
Whereas $E_0$ has positive value and we can write
\begin{equation} \label{E0viau01}
E_0 = \sqrt{1 - \frac{2}{3} M u_0} \, ,
\end{equation}
the value of $x_0$ can have both signs. From expression for $x_0$ in (\ref{xE}), we see that the second term in numerator is always bigger (or equal) than the first term, which means that the sign of $x_0$ is determined by sign in front of the second term in numerator. Therefore $x_0>0$ stands for co-rotating orbits and $x_0<0$ for counter-rotating orbits. We can write:
\begin{equation} \label{x0viau0}
x_0 = \mp \frac{1}{\sqrt{3} u_0} \, .
\end{equation}
For $E_0$ we can also write:
\begin{equation} \label{E0viau02}
E_0 = \frac{6Mu_0 - 3u_0^2 a^2 - 1}{6au_0^2 x_0} = \mp \frac{(6Mu_0 - 3u_0^2 a^2 - 1) \sqrt{3}}{6au_0} .
\end{equation}

Three equations for small-spin corrections $x_1$, $E_1$, $u_1$, partially simplified with using the third equation in (\ref{system-null}) and relation $x_0^2 u_0^2 = 1/3$, are:
\begin{equation}
\left\{
\begin{aligned}
& (- 6 x_0 - 6aE_0 + 18 M x_0 u_0) \, x_1 - 6a x_0 \, E_1 + \\
& + 6 a E_0^2 + 6E_0 x_0 - 27 E_0 M x_0 u_0 + 5aMu_0 = 0 \, ,\\
& 2 M x_0 u_0^3 \,x_1 - 2 E_0 \, E_1 + M a u_0^3  - 3 E_0 M x_0 u_0^3 =  0 \, ,\\
& (-3 x_0 - 3a E_0 + 18 M x_0 u_0) \, x_1 - 3a x_0 \, E_1 + 9 M x_0^2 \, u_1 + \\
&  + 3a E_0^2 + 3 E_0 x_0 - 27 E_0 M x_0 u_0 + 10aM u_0  =  0 \, .\\
\end{aligned}
\right.
\label{system-linear}
\end{equation}

After significant simplifications with using expressions (\ref{x0E0viau0}), (\ref{x0viau0}), (\ref{E0viau02}) we succeed to obtain the corrections in a simple form:
\begin{equation}
\begin{aligned}
E_1 &= -x_0 M u_0^3 = \pm \frac{1}{\sqrt{3}} M u_0^2, \\
x_1 &= - \frac{(1-6Mu_0 + 9a^2 u_0^2) x_0}{4a} = \pm \frac{1-6Mu_0 + 9a^2 u_0^2}{4\sqrt{3} a u_0} , \\
u_1 &= - \frac{u_0 (1-6Mu_0 + 5a^2 u_0^2)}{2a}. \\
\end{aligned}
\label{lineinye-popravki}
\end{equation}
To avoid singularity in the Schwarzschild case ($a=0$, $u_0=1/6M$), we can also rewrite (\ref{lineinye-popravki}) as
\begin{equation}
\begin{aligned}
E_1 &= -x_0 M u_0^3 = \pm \frac{1}{\sqrt{3}} M u_0^2, \\
x_1 &= \pm \frac{1}{\sqrt{3}} (3a u_0 \pm 2 \sqrt{Mu_0} ) = \frac{1}{\sqrt{3}} (2 \sqrt{Mu_0} \pm   3a u_0 ) , \\
u_1 &= -4 u_0^2 (a u_0 \pm \sqrt{M u_0}). \\
\end{aligned}
\label{lineinye-popravki-schw}
\end{equation}

Let us summarise the results of this section. To obtain ISCO parameters for given $a$ in the form (\ref{podstanovka}) one should find $u_0$ from equation
\begin{equation}
1- 3 a^2 u_0^2 - 6 M u_0 \mp 8 a \sqrt{Mu_0^3} =0 ,
\label{eq-u0}
\end{equation}
and then calculate $x_0$ with using (\ref{x0viau0}), $E_0$ with using (\ref{E0viau01}), $x_1$, $E_1$, $u_1$ with using (\ref{lineinye-popravki}) or (\ref{lineinye-popravki-schw}). The value $J$ can be found by using $J=x+aE$:
\begin{equation}
J = J_0 + s J_1 = (x_0 + aE_0) + s(x_1 + aE_1).
\end{equation}
The ISCO radius $r$ can be found by using $r=1/u$:
\begin{equation}
r = r_0 + s r_1 = \frac{1}{u_0}  - s \frac{u_1}{u_0^2} .
\end{equation}

We calculate also the ISCO frequency for arbitrary value of $a$, as described at the end of section \ref{section-EPspin}. Using (\ref{x0viau0}), (\ref{E0viau02}) and (\ref{lineinye-popravki}) in (\ref{Omega-definition}) we obtain $\Omega$ as
\begin{equation}
\Omega = \Omega_0 + s \, \Omega_1,
\end{equation}
\begin{equation} \label{Omega-arbitrary-a}
\Omega_0 = \frac{\sqrt{M} u_0^{3/2}}{a \sqrt{M} u_0^{3/2} \mp 1} , \;\; \Omega_1 = \frac{9 \sqrt{M} u_0^3 (\sqrt{M} \pm a\sqrt{u_0}) }{2 \left(1 \mp a \sqrt{M} u_0^{3/2} \right)^2} ,
\end{equation}
or, in terms of $r_0$ as
\begin{equation}
\Omega_0 = \frac{\sqrt{M}}{a\sqrt{M} \mp  r_0^{3/2}   } ,  \;\;
\Omega_1 = \frac{9 \sqrt{M} ( \sqrt{r_0 M} \pm a ) }{2 \sqrt{r_0} \left(r_0^{3/2} \mp a \sqrt{M} \right)^2 } .
\end{equation}

In all the formulae the upper sign corresponds to the antiparallel orientation of particle's angular momentum $\mathbf{J}$ and BH spin $\mathbf{a}$ (counter-rotation, $J<0$), the lower -- to the parallel one (corotation, $J>0$); $s$ is the projection of spin on the $z$-axis and can be positive (spins of particle and BH are parallel) or negative (antiparallel); $a$ is positive or equal to zero; the $z$-axis is chosen to be parallel to BH spin $\mathbf{a}$.

\section{Small-spin corrections for ISCO parameters in Schwarzschild metric} \label{section-schw}

In order to illustrate the obtained results, let us consider some important limit cases: Schwarzschild metric and Kerr metric, when the BH rotation is slow ($a \ll M$) and extremely fast ($a \to M$). The latter case is called "extreme Kerr BH", for at the values of $a$ exceeding $M$ the BH disappears and we have a naked singularity \cite{Chandra} without an event horizon.

In the Schwarzschild case there is no prefential direction associated with BH spin ($a=0$), so $z$-direction can be chosen arbitrarily. Choosing $z$-axis along $\mathbf{J}$ vector, and taking $u_0 = 1/6M$, we obtain:
\begin{equation}
\begin{aligned}
J_{\mathrm{\, ISCO}}&=2 \sqrt{3} M + \frac{\sqrt{2}}{3}s_J ,\\
E_{\mathrm{\, ISCO}}&= \frac{2 \sqrt{2}}{3} -\frac{1}{36 \sqrt{3}}\frac{s_J}{M} ,\\
r_{\mathrm{\, ISCO}}&= 6M -2\sqrt{\frac{2}{3}}s_J ,\\
\Omega_{\mathrm{\, ISCO}}&= \frac{1}{6 \sqrt{6}M} +\frac{s_J}{48 M^2} .\\
\end{aligned}
\label{Schw}
\end{equation}
Here instead of $s$, which is the projection on $z$-axis that does not unabiguously correspond to any physical direction in Schwarzschild case, we use $s_J$ which is the projection of particle's spin upon the direction of $\mathbf{J}$ and is positive when the particle's spin is parallel to it and negative when it is antiparallel. Value $J$ is considered as positive in this case.

Small-spin corrections for Schwarzschild metric were derived by Favata \cite{Favata2011}.

\section{Corrections for ISCO parameters for slowly rotating BH} \label{section-slowly}

The case of slowly rotating Kerr BH ($a \ll M$) allows us to separate corrections that arise from different types of angular momentum coupling.

Our general formulae, presented in section \ref{section-arbitrary}, allow us to obtain correction terms proportional to $a$, $s$ and $as$. Considering $a \ll M$ in equation (\ref{eq-u0}), we get $u_0$:
\begin{equation}
u_0 = \frac{1}{6M} \mp \frac{1}{9M}\sqrt{\frac{2}{3}} \frac{a}{M} +\frac{13}{216M}\frac{a^2}{M^2}.
\label{u0-slow-Kerr}
\end{equation}
Using the first two terms from (\ref{u0-slow-Kerr}) in (\ref{E0viau01}), (\ref{x0viau0}), (\ref{lineinye-popravki-schw}), we obtain $x_0$, $E_0$, $x_1$, $E_1$ and $u_1$. Note that when using formulae (\ref{lineinye-popravki}) for calculation of corrections $E_1$ and $u_1$, the quadratic terms from (\ref{u0-slow-Kerr}) should be also taken into account. Corrections $a$, $s$, $as$ for ISCO parameters can be as well obtained directly by solving the system (\ref{p}), with keeping terms linear in $a$ and $s$. 

If we don't assume any ratio between values $a$ and $s$, a full description of the problem requires also to obtain quadratic terms $s^2$ and $a^2$ in ISCO parameters. The scheme for obtaining the squared corrections is analogous to the linearisation procedure, only we need to express the parameters in the form
\begin{equation}
\begin{aligned}
x &= x_0 + s x_1 + s^2 x_2,\\
E &= E_0 + s E_1 + s^2 E_2,\\
u &= u_0 + s u_1 + s^2 u_2,\\
\end{aligned}
\label{} \end{equation}
and retain the squared in spin terms in the system (\ref{p}).

For our purposes we can use more simple way: corrections $s^2$ can be found from solution of the system (\ref{p}) for Schwarzschild case, and $a^2$-corrections can be found from consideration of the spinless case.

\begin{figure}
\centerline{\hbox{\includegraphics[width=0.5\textwidth]{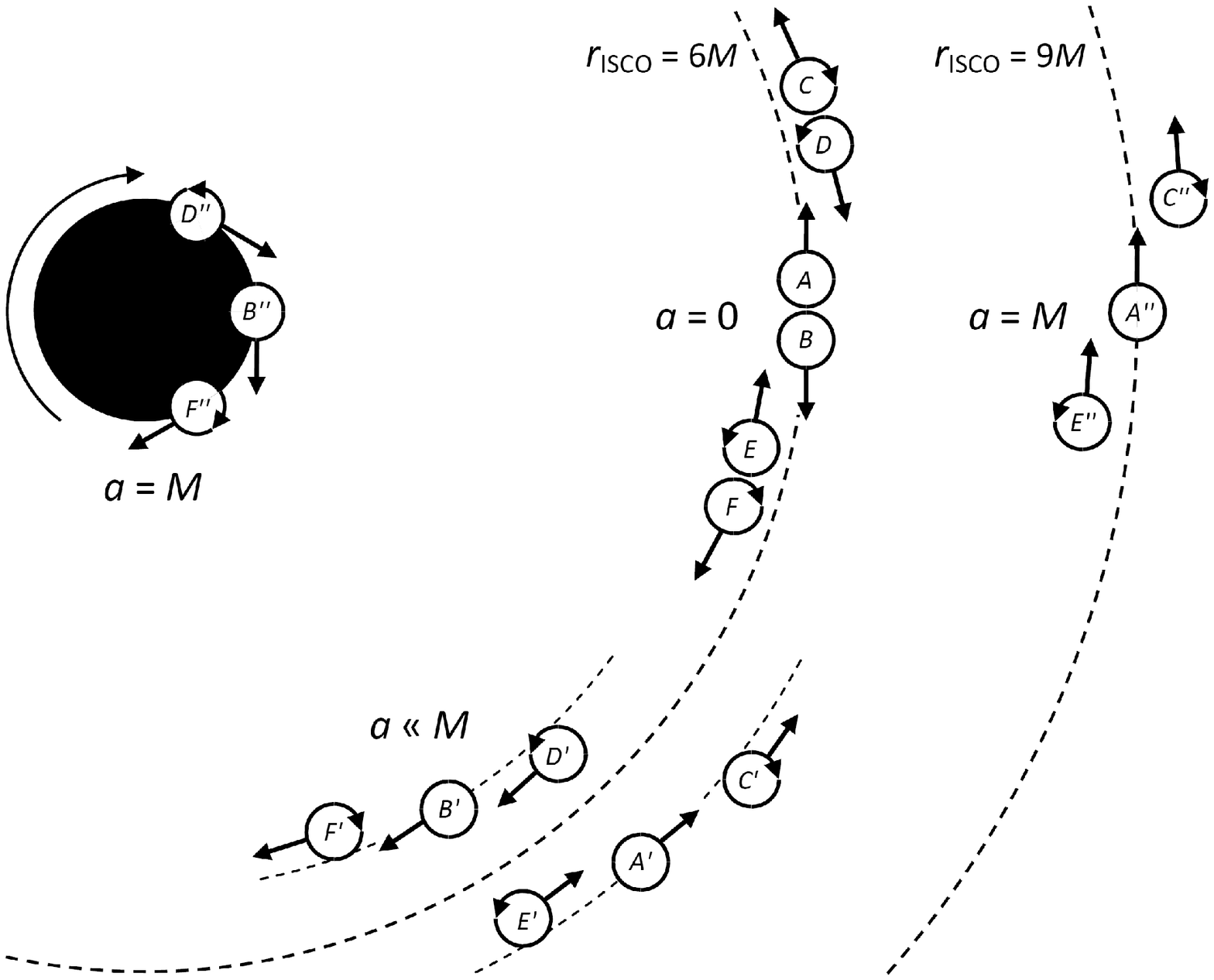}}}
\caption{Radii of ISCO for different directions of motion and spin of test particles. \\
Six particles, $A$, $B$, $C$, $D$, $E$, $F$, with different directions of spin and orbital motion are presented, for different Kerr parameter $a$. For $a=0$ we use just letters (e.g., $A$), for $a \ll M$ we use primes (e.g., $A'$), for $a=M$ we use double primes (e.g., $A''$). Direction of particle's spin is shown by an arrow on a circle: clockwise arrow ($C$ and $F$) indicates particles co-rotating with black hole, anticlockwise arrow ($D$ and $E$) indicates counter-rotating particles, for spinless particles ($A$ and $B$) we use circles without arrow. Direction of orbital motion around BH is shown by straight arrow outside circles. Size of BH is shown for the case of the extreme Kerr space-time. The ISCO radii for $a \ll M$ are drawn on basis of only the first three terms in formula (\ref{Kerr-slow}), with $s \ll a \ll M$. \\
In the case of a Schwarzschild BH ($a=0$) the ISCO radius for spinless particles equals to $6M$, independently of direction of orbital motion, see $A$ and $B$. In the case of a Kerr BH the ISCO radius depends on direction of orbital motion. For slowly rotating BH, $a \ll M$: counter-rotating particle $A'$ has orbit with $r_{\mathrm{\, ISCO}}>6M$, co-rotating particle $B'$ has orbit with $r_{\mathrm{\, ISCO}}<6M$. For the extreme Kerr BH, $a=M$, a counter-rotating particle $A''$ has orbit with $r_{\mathrm{\, ISCO}}=9M$, a co-rotating particle $B''$ has orbit with $r_{\mathrm{\, ISCO}}=M$. \\
Presence of spin increases or decreases ISCO radius in comparison with spinless case, see particles $C$, $D$, $E$, $F$ (also $C'$, $C''$ and so on). Remind that $z$-axis is directed along the axis of rotation of BH, so $a>0$. Total angular momentum $J=J_z$: $J>0$ corresponds to co-rotating of orbit, $J<0$ corresponds to counter-rotating orbit. Spin $s>0$ corresponds to spin vector parallel to $a$ (corotation with BH), spin $s<0$ is antiparallel to it. The ISCO radius of spinning particle in the Schwarzschild case depends on $s_J$, projection of spin vector on total angular momentum vector. Therefore particles $C$ and $D$ (also $E$ and $F$), which have different ISCO radii in the Kerr metric, have the same ISCO radius in Schwarzschild case.
For the extreme Kerr BH particles $B''$, $D''$, $F''$ have the same ISCO radii.}
\label{figure-Im2}
\end{figure}

Finally, we obtain:
\begin{equation}
\begin{aligned}
J_{\mathrm{\, ISCO}}&= \mp 2\sqrt{3}M -\frac{2 \sqrt{2}}{3}a +\frac{\sqrt{2}}{3}s \pm\frac{11}{36 \sqrt{3}}\frac{a}{M}s \pm\\&\pm \frac{4 \sqrt{3} M}{27}\left( \frac{a}{M} \right)^2 \pm \frac{1}{4 M \sqrt{3}}s^2,\\
E_{\mathrm{\, ISCO}}&= \frac{2 \sqrt{2}}{3} \pm\frac{1}{18 \sqrt{3}}\frac{a}{M} \pm \frac{1}{36 \sqrt{3}}\frac{s}{M} -\frac{\sqrt{2}}{81}\frac{a}{M}\frac{s}{M} -\\&-\frac{5}{162 \sqrt{2}}\left( \frac{a}{M} \right)^2 -\frac{5}{432 \sqrt{2}M^2}s^2 ,\\
r_{\mathrm{\, ISCO}}&= 6 M \pm 4 \sqrt{\frac{2}{3}}a \pm 2 \sqrt{\frac{2}{3}}s + \frac{2}{9} \frac{a}{M}s -\\&-\frac{7M}{18}\left( \frac{a}{M} \right)^2 -\frac{29}{72M}s^2 .\\
\end{aligned}
\label{Kerr-slow}
\end{equation}
Different situations of mutual orientation of BH and particle spin are presented on Figure \ref{figure-Im2}, where the case $a \ll M$ is drawn on the basis of the first three terms in formula (\ref{Kerr-slow}) for $r_{\mathrm{\, ISCO}}$. For example, for particle $C'$ we should use the upper sign and $s>0$, for particle $D'$ we should use the lower sign and $s<0$.

This result allows to understand the character of the spin-orbital coupling in GR. From the terms $\pm 4 \sqrt{2/3}a$ and $\pm 2 \sqrt{2/3}s$, we see that the the spin-orbital coupling depends on mutual orientation of the spin and angular momentum vectors: it is attractive and the radius decreases, when they are parallel, and repulsive, if they are antiparallel.

The terms with $as$, $a^2$, $s^2$ in formulae (\ref{Kerr-slow}) are of second infinitesimal order and include information about interactions of particle spin, BH spin and orbital momentum. Considering the formula for radius it is easy to show that the trinomial containing these three terms is negative for all possible values of $a$ and $s$. Therefore, quadratic corrections always lead to decrease of the ISCO radius.

Substituting values of $J_{\mathrm{\, ISCO}}$, $E_{\mathrm{\, ISCO}}$ and $r_{\mathrm{\, ISCO}}$ from (\ref{Kerr-slow}) into (\ref{Omega-definition}), we obtain the ISCO frequency as
\begin{equation}
\begin{aligned}
& \Omega_{\mathrm{\, ISCO}} = \mp \frac{1}{6 \sqrt{6}M} +\frac{11}{216 M}\frac{a}{M} +\frac{1}{48 M^2}s \, \mp\\ 
&\mp \left( \frac{1}{18 \sqrt{6}M}\frac{as}{M^2} + \frac{59}{648 \sqrt{6}M} \frac{a^2}{M^2} + \frac{97}{3456 \sqrt{6}M}\frac{s^2}{M^2} \right).
\end{aligned}
\end{equation}

\section{Small-spin linear corrections for extreme Kerr BH} \label{section-extreme-Kerr}

\begin{figure}
\centerline{\hbox{\includegraphics[width=0.45\textwidth]{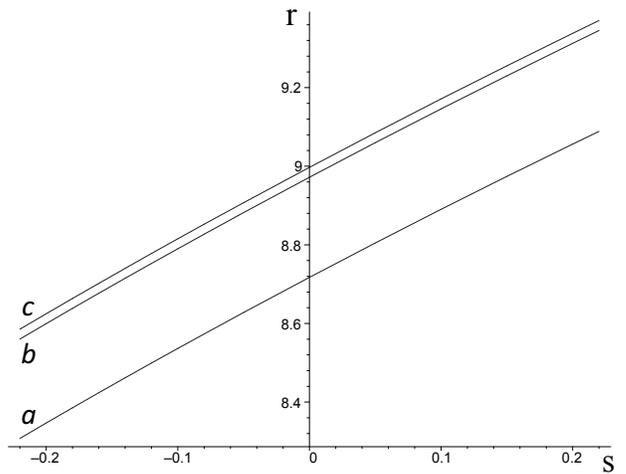}}}
\caption{The ISCO radii as a function of particle's spin for different values of $a$ close to $M$: $a/M = $ 0.9 (a), 0.99 (b), 0.999 (c), in case of counter-rotation. In limit $a \rightarrow M$ the ISCO radius at $s=0$ tends to the $9M$. }
\label{figure-eKerr-rr}
\end{figure}

In the extreme Kerr background ($a \to M$) for a spinless particle the ISCO radius is $9M$ in the antiparallel case (counter-rotation), and $M$ in the parallel case (co-rotation). Numerical solution of the system (\ref{p}) for $a$ close to $M$ shows that ISCO radius has very different behaviour for co-rotating ($J>0$) and counter-rotating ($J<0$) particles, see Fig. \ref{figure-eKerr-co} and \ref{figure-eKerr-rr}. We will therefore consider these cases separately.

In the antiparallel case we get, using (\ref{E0viau01}), (\ref{x0viau0}), (\ref{lineinye-popravki}), (\ref{Omega-arbitrary-a}): 
\begin{equation}
\begin{aligned}
J_{\mathrm{\, ISCO}}&= -\frac{22 \sqrt{3}}{9}M +\frac{82\sqrt{3}}{243}s,\\
E_{\mathrm{\, ISCO}}&=\frac{5 \sqrt{3}}{9} +\frac{\sqrt{3}}{243}\frac{s}{M},\\
r_{\mathrm{\, ISCO}}&= 9M +\frac{16}{9}s,\\
\Omega_{\mathrm{\, ISCO}} &= -\frac{1}{26M} +\frac{3s}{338 M^2}.\\
\end{aligned}
\label{Kerr-counter}
\end{equation}

For the parallel case we search for ISCO parameters at values of $a=(1-\delta)M$, $\delta \ll 1$, because divergences may appear at $a=1$ at some ways of calculations. Using (\ref{E0viau01}), (\ref{x0viau0}), (\ref{lineinye-popravki}), (\ref{Omega-arbitrary-a}), we get:
\begin{equation}
\begin{aligned}
J_{\mathrm{\, ISCO}}&=\left( \frac{2}{\sqrt{3}} +\frac{2 \times 2^{2/3} \delta^{1/3}}{\sqrt{3}} \right)M +\\
&+\left( -\frac{2}{\sqrt{3}} +\frac{4 \times 2^{2/3}\delta^{1/3}}{\sqrt{3}} \right)s,\\
E_{\mathrm{\, ISCO}}&=\left( \frac{1}{\sqrt{3}} +\frac{2^{2/3} \delta^{1/3}}{\sqrt{3}} \right) +\\
&+\left( -\frac{1}{\sqrt{3}} +\frac{2 \times 2^{2/3}\delta^{1/3}}{\sqrt{3}} \right) \frac{s}{M},\\
r_{\mathrm{\, ISCO}}&= \left( 1 +2^{2/3}\delta^{1/3} \right)M -2 \times 2^{2/3} \delta^{1/3}s,\\
\Omega_{\mathrm{\, ISCO}} &= \frac{1}{2M}  - \frac{3 \times 2^{2/3} \delta^{1/3}}{8M} + \frac{9 \times 2^{2/3} \delta^{1/3}}{16M^2} s .\\
\end{aligned}
\label{Kerr-co}
\end{equation}
We see that in the case of $a=M$ ($\delta=0$) the corrections, linear in spin, are absent in formulae for ISCO radius and frequency. This was also demonstrated in \cite{Abramowicz1979}. In the next section we discuss this feature in details and find the exact in spin expressions for the ISCO parameters.

The radius of horizon of the Kerr BH,
\begin{equation}
r_{\mathrm{\, hor}} = M + \sqrt{M^2-a^2},
\end{equation}
in the case $a=(1-\delta) M$ will have the form \cite{Bardeen1972, LL2}
\begin{equation}
r_{\mathrm{\, hor}} = M (1 + \sqrt{2 \delta}) \, .
\end{equation}
As $s \ll M$ and $\delta^{1/2} < \delta^{1/3}$ for $\delta \ll 1$, it will always be that $r_{\mathrm{\, ISCO}} > r_{\mathrm{\, hor}}$.

\section{Exact solution for extreme Kerr BH in the case of co-rotation}
\label{section-exact}

\begin{figure}
\centerline{\hbox{\includegraphics[width=0.45\textwidth]{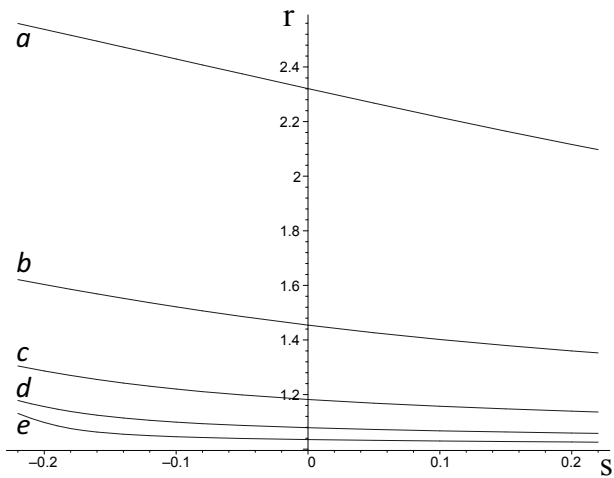}}}
\caption{The ISCO radius as a function of particle's spin for different values of $a$ close to $M$: $a/M = $ 0.9 (a), 0.99 (b), 0.999 (c), 0.9999 (d), 0.99999 (e), in case of co-rotation. In limit $a \rightarrow M$ curves tend to the horizontal asymptot $r_{\mathrm{\, ISCO}}=1$.}
\label{figure-eKerr-co}
\end{figure}

In the work \cite{TanakaMino1996} on basis of the numerical calculation, it was noticed that in the extreme Kerr background for the parallel case the magnitude of test-body's spin does not influence the radius of the last stable orbit and it always remains equal to $M$. Solving numerically Eqs (\ref{p}), we plot Fig. \ref{figure-eKerr-co} where this conclusion is distinctly visible. We also succeeded in proving this analytically.

On basis of results of the numerical solution we have substituted $u=1/M$ into the system (\ref{p}) and have obtained the expressions for $x$ and $E$, which is the analytical proof of existence of such solution of the system. The exact solution is then
\begin{equation}
\begin{aligned}
J_{\mathrm{\, ISCO}}&= 2 \frac{M^2-s^2}{M\sqrt{3 +6s/M}},\\
E_{\mathrm{\, ISCO}}&= \frac{M^2-s^2}{M^2\sqrt{3 +6s/M}},\\
r_{\mathrm{\, ISCO}}&=M.\\
\end{aligned}
\label{Kerr-exact}
\end{equation}
Using (\ref{Kerr-exact}) in (\ref{Omega-definition}), we obtain the ISCO frequency as
\begin{equation}
\Omega_{\mathrm{\, ISCO}} = \frac{1}{2M}  .
\end{equation}

Developing (\ref{Kerr-exact}) in series in $s$, we get the linear in spin terms which are in agreement with (\ref{Kerr-co}) at $\delta=0$. This result allows us to obtain the full in spin correction to the nearly extreme black hole, with $a=(1-\delta) M$, considered in the section \ref{section-extreme-Kerr}. For this purpose after substituting this value for $a$ we develop the system (\ref{p}) in series in $\delta$ and seek for the solution in the form $y=y_0(s) +y_1(s) \delta^{1/3}$, where $y$ stands for $x,E,r$. After some algebraic calculation we obtain for the nearly extreme Kerr metric the expressions:
\begin{widetext}
\begin{equation}
\begin{aligned}
J_{\mathrm{\, ISCO}}&= 2 M E_{\mathrm{\, ISCO}} \, ,\\
E_{\mathrm{\, ISCO}}&= \frac{M^2-s^2}{M^2\sqrt{3 +6s/M}} +\frac{(M^2 -s^2)^{1/3}(2M +s)^{2/3} Z(M,s)^{2/3}}{\sqrt{3}M^{5/2}(M +2s)^{3/2}}  \, \delta^{1/3},\\
r_{\mathrm{\, ISCO}}&=M +\frac{M(M^2 -s^2)^{1/3}(2M +s)^{2/3}}{Z(M,s)^{1/3}}  \,  \delta^{1/3}, \\
\Omega_{\mathrm{\, ISCO}} &= \frac{1}{2M} -\frac{3(M -s)^{1/3}(M +2s)}{4(2M +s)^{1/3}(M +s)^{2/3} Z(M,s)^{1/3}}\,  \delta^{1/3} ,\\
\mbox{where} \;\; Z(M,s) &\equiv M^4 +7M^3s +9M^2 s^2 +11M s^3 -s^4 .\\
\end{aligned}
\label{Kerr-delta}
\end{equation}
\end{widetext}
Developing this result in series in $s$, one obtains all terms in expressions (\ref{Kerr-co}).

Note that derivation of MPD equations was performed under condition $s \ll M$, so always we have $J_{\mathrm{\, ISCO}} > 0$, $E_{\mathrm{\, ISCO}} > 0$ in (\ref{Kerr-exact}) and (\ref{Kerr-delta}). Indeed, equations of motion of a spinning particle were derived under the assumption that its characteristic size is very small compared with the characteristic length of the background field (e.g. the distance from the central body) \cite{Papapetrou1951a}, \cite{SaijoMaeda1998}. M{\o}ller \cite{Moller1949} had shown in the context of special relativity that the classical spinning body must have a certain minimum characteristic size connected with its specific spin, see also Wald \cite{Wald1972}. For particle at ISCO, in our system of units the above conditions reduce to $s \ll M$, see also \cite{Rasband1973}, \cite{SaijoMaeda1998}, \cite{Apostolatos1996}.

\section{Spin corrections for arbitrary circular orbits}
\label{section-circular}

In previous parts of the present paper we have discussed ISCO, and investigated how the presence of spin influences parameters of ISCO for different given $a$. Let us now discuss not innermost stable but arbitrary circular orbits. We will investigate how the presence of spin influences orbit of given radius -- namely, how $E$, $J$, $\Omega$ of a circular orbit of given radius are changed if a test particle has spin.

Let us suppose that inverse radius $u$ of circular orbit is known and consider the system (\ref{n}). We shall seek for its solution in the form
\begin{equation}
\begin{aligned}
x &=x_0+s x_1,\\
E &=E_0+ s E_1.\\
\end{aligned}
\end{equation}
Here $x_0$ and $E_0$ are the values of $x$ and $E$ in the "spinless" case, see (\ref{xE}), $x_1$ and $E_1$ are the linear corrections associated with spin. We obtain for the linear corrections:
\begin{widetext}
\begin{equation} \label{circ-first}
\begin{aligned}
x_1 (u) &=\frac{2 E_0^2 x_0 - 9 M E_0^2 x_0 u - 3 a^2 M x_0^3 u^5+ 2 a E_0^3+ 8 a M E_0 x_0^2 u^3}{2(a E_0^2+ E_0 x_0 - 3 M E_0 x_0 u+ a M x_0^2 u^3)},\\
E_1 (u) &=\frac{M u^3 x_0 (-E_0 x_0 +3 a^2 E_0 x_0 u^2 -a E_0^2 -4 a M x_0^2 u^3 +3 a x_0^2 u^2)}{2(a E_0^2+ E_0 x_0 - 3 M E_0 x_0 u+ a M x_0^2 u^3)}.\\
\end{aligned}
\end{equation}
\end{widetext}
These expressions are valid for both stable and unstable circular orbits with the zero order ISCO parameters given by (\ref{xE}). Value of total angular momentum $J=J_0 + s J_1$ can be obtained using $J=x+aE$.

The expressions (\ref{circ-first}) can be simplified further. Using (\ref{xE}) we obtain:
\begin{widetext}
\begin{equation} \label{E1-long}
E_1 =\frac{M u^{5/2} (  a\sqrt{u} \pm \sqrt{M} ) ( 1+3a^2u^2 \pm 4a\sqrt{Mu^3} ) }{2 (1 - 3 M u \mp 2 a \sqrt{Mu^3} )^{3/2}}
\end{equation}
\begin{equation} \label{J1-long}
J_1 =\frac{ 3Ma^4u^5 \pm 7a^3M^{3/2} u^{9/2} + 4M^2 u^4 a^2 \mp 3\sqrt{M} u^{7/2} a^3 + 2Mu^3a^2 \pm 21 M^{3/2} u^{5/2} a + 18 M^2 u^2 \mp 9 a \sqrt{Mu^3}  -13Mu +2   }{2 (1 - 3 M u \mp 2 a \sqrt{Mu^3} )^{3/2} }
\end{equation}
\end{widetext}
Written as functions of $r$ the resulting formulae are
\begin{equation}
\begin{aligned}
E_0 &=\frac{r^{3/2} -2Mr^{1/2} \mp aM^{1/2}}{r^{3/4}(r^{3/2} -3Mr^{1/2} \mp 2aM^{1/2})^{1/2}},\\
J_0 &=\frac{\mp M^{1/2}(r^2 \pm 2aM^{1/2}r^{1/2} +a^2)}{r^{3/4}(r^{3/2} -3Mr^{1/2} \mp 2aM^{1/2})^{1/2}}.\\
\end{aligned}
\end{equation}

\begin{widetext}
\begin{equation}
\begin{aligned}
E_1 &=\frac{M (a \pm \sqrt{Mr} )(r^2 + 3a^2  \pm 4a \sqrt{Mr} )}{2r^{11/4}(r^{3/2} -3M \sqrt{r} \mp 2a \sqrt{M} )^{3/2}},\\
J_1 &=\frac{2r^5 -13Mr^4 \mp 9aM^{1/2}r^{7/2} +18M^2r^3 \pm 21aM^{3/2}r^{5/2} +2a^2Mr^2 \mp 3a^3 M^{1/2}r^{3/2} +4a^2M^2r \pm 7a^3 M^{3/2}r^{1/2} +3a^4M}{2r^{11/4}(r^{3/2} -3M \sqrt{r} \mp 2a \sqrt{M})^{3/2}}.\\
\end{aligned}
\end{equation}
\end{widetext}

Substituting (\ref{E1-long}) and (\ref{J1-long}) into (\ref{Omega-definition}), we obtain angular frequency $\Omega = \Omega_0 + s \Omega_1$ at the circular orbit with a given inverse radius $u$ as
\begin{equation}
\begin{aligned}
\Omega_0 &=\frac{\sqrt{Mu^3}}{ a \sqrt{Mu^3} \mp 1} , \\
\Omega_1 &= - \frac{3 u^3 \sqrt{M} (\sqrt{M} \pm a \sqrt{u} )}{2 ( a \sqrt{Mu^3} \mp 1 )^2} , \\
\end{aligned}
\end{equation}
or, as function of radial coordinate $r$, as
\begin{equation}
\begin{aligned}
\Omega_0 &= \frac{\sqrt{M}}{a \sqrt{M} \mp r^{3/2}} , \\
\Omega_1 &= -\frac{3 \sqrt{M} (\sqrt{Mr} \pm a)}{2 \sqrt{r} (a\sqrt{M} \mp r^{3/2})^2} . \\
\end{aligned}
\end{equation}

The linear corrections for the energy, angular momentum and frequency at a given radius for the circular orbits are given also in \cite{Abramowicz1979}, \cite{Calvani1980} in another presentation, see also Appendices A and B in \cite{Steinhoff2012} where comparison of spin orientation in different papers is discussed.

\section{Conclusions}
\label{section-conclusions}

(i) The linear in spin corrections for the ISCO parameters: radius, total angular momentum, energy, orbital angular frequency for arbitrary Kerr parameter $a$, are found analytically, see Section \ref{section-arbitrary}.

(ii) Small-spin linear corrections for Schwarzschild metric (firstly obtained by Favata \cite{Favata2011}) were derived, see formulae (\ref{Schw}).

(iii) Expressions for the ISCO parameters of a slowly-rotating Kerr BH up to second order in spin and Kerr parameter are obtained, see (\ref{Kerr-slow}). The case of a slowly rotating Kerr BH gives the idea of spin-orbital coupling's influence on the parameters of ISCO. It is found that spin-orbital coupling has an attractive character when the total angular momentum $J$ and spin of either BH or a test-body are parallel and repulsive when they are antiparallel.

(iv) Linear in spin corrections to the ISCO parameters for extreme Kerr BH both for counter- and co-rotating cases are obtained, see (\ref{Kerr-counter}) and (\ref{Kerr-co}).

(v) The exact parameters of ISCO for a spinning test-body in the extreme Kerr background in corotation case are found, see (\ref{Kerr-exact}) and (\ref{Kerr-delta}). It is proved analytically that radius and angular frequency of such orbits are independent of the particle's spin while the values of energy and total angular momentum depend on it.

(vi) Figures \ref{figure-Im1} and \ref{figure-Im2} illustrate the number of possible configuration for rotating BH and a spinning test particle with parameters calculated in the present work.

\section*{Acknowledgments}

We are greatful to an anonymous referee for useful suggestions leading to improvement of the paper.

Authors would like to thank I.D. Novikov and V. Perlick for valuable comments during discussions of results of the research. 

The work of GSBK, OYuT and PIJe was partially supported by the Russian Foundation for Basic Research Grant No. 14-02-00728 and the Russian Federation President Grant for Support of Leading Scientific Schools, Grant No. NSh-261.2014.2.

The work of GSBK was partially supported by the Russian Foundation for Basic Research Grant No. OFI-M 14-29-06045.

The work of OYuT and PIJe was partially supported by the Russian Federation President Grant for Support of Young Scientists, Grant No. MK-2918.2013.2.

\bibliographystyle{ieeetr}

\end{document}